\documentclass[cits]{PoS}

\usepackage{epsfig}

\renewcommand{\i}{\mathrm{i}}
\newcommand{\Tr}{\mathrm{Tr}}

\newcommand{\Dsl}{\rlap{\,/}{{D}}}

\newcommand{\lk}{\left<}
\newcommand{\ls}{\left|}

\newcommand{\rk}{\right>}
\newcommand{\rs}{\right|}
\newcommand{\Z}{\mathrm{Z}}

\title{The Polyakov Loop and the Eigenvalues of the\\ Dirac Operator}

\ShortTitle{The Polyakov Loop and the Eigenvalues of the Dirac
  Operator}

\author{\speaker{Wolfgang S\"oldner}\thanks{This work has been supported in
    part by contract DE-AC02-98CH10886 with the U.S. Department of Energy}\\
  Physics Department, Brookhaven National Laboratory, Upton, NY 11973, USA\\
  E-mail: \email{soeldner@bnl.gov}}

      \abstract{Aiming at the link between confinement and chiral
        symmetry the Polyakov loop represented as a spectral sum of
        eigenvalues of the Dirac operator was subject of recent
        studies. We analyze the volume dependence as well as the
        continuum behavior of this quantity for quenched QCD using
        staggered fermions. Furthermore, we present first results
        using dynamical configurations.  }

\FullConference{The XXV International Symposium on Lattice Field Theory\\
		 July 30-4 August 2007\\
		 Regensburg, Germany}

\begin{document}

\section{Motivation}

Lattice simulations suggest that the chiral phase transition and
deconfinement phase transition appear at the same temperature. It is
believed that there is a connection between both phase transitions.
While for the chiral phase transition we have a well established
picture of the symmetry breaking mechanism the picture of the
deconfinement phase transition remains unclear. Although there has
been much progress in the last years the final link connecting both
phase transitions is still missing.

The order parameter of the chiral phase transition is the chiral
condensate $\langle \bar q q \rangle$. Banks and Casher~\cite{BC}
related the chiral condensate to the eigenvalue density $\rho$ of the
Dirac operator near zero,
\begin{equation}
  \langle \bar q q \rangle = - \pi \rho(\lambda=0).
\end{equation}
Recently, Gattringer~\cite{Gattringer:2006ci} established a formula
which relates the eigenvalues of the Dirac operator to the Polyakov
loop $P$, the order parameter of the deconfinement phase transition in
the quenched approximation. This relation provides a natural link
between the chiral condensate and the Polyakov loop via the
eigenvalues of the Dirac operator. The hope is to obtain some insight
into how both phase transitions are connected.

After a short introduction we will discuss several aspects of this new
relation between the Polyakov loop and the eigenvalues, in particular
we focus on the volume scaling and the continuum limit. We will
present numerical results for both quenched and dynamical QCD and will
also compare to the free case.

\section{Introduction}
Starting point of our discussion are the eigenvalues $\lambda$ of the
massless staggered Dirac operator, $\Dsl \Psi = \pm \i \lambda \Psi$ with
$\lambda real > 0$. The massless staggered Dirac operator is defined by
\begin{equation}
  \Dsl_{xy} =\frac{1}{2} \sum_{\mu=1}^4 \bigg[\eta_{x\mu}
  U_{x\mu}^\dagger \cdot \delta_{(x+\hat{\mu}),y} - \eta_{(x-\hat{\mu})\mu}
  U_{(x-\hat{\mu})\mu} \cdot \delta_{(x-\hat{\mu}),y} \bigg],
\end{equation}
where $\eta_{x\mu}$ is the usual staggered phase factor and $U_{x\mu}$
are the link variables. Note that we use periodic boundary conditions
in all four directions for the calculation of the eigenvalues. The
Polyakov loop is defined by $P=\frac{1}{N_c \, N_s^3} \sum_{\vec{n}}
\Tr_c \left[ \prod_{n_4=1}^{N_t} U_4(\vec{n},n_4) \right]$ and can be
expressed in terms of the eigenvalues in the following
way~\cite{Gattringer:2006ci},
\begin{equation}
  P = \i^{N_t} \frac{2 \, 2^{N_t}}{3\, N_t\, N_s^3} 
  \sum_i \left\{1 \cdot \lambda_{i,1}^{N_t} +
    z  \cdot \lambda_{i,z^\ast}^{N_t} + z^\ast  \cdot \lambda_{i,z}^{N_t}
  \right\}.
\label{eq:P}
\end{equation}
$N_s$ and $N_t$ is the spatial and temporal extension, respectively,
and $\Z_3 = \{1,z,z^\ast \}$. Note that $N_t$ has to be even for
staggered fermions. The sum over $i$ in Eq.~\ref{eq:P} is meant to sum
over all eigenvalues $\lambda_{i,X}$, where $\lambda_{i,X}$ stands for
the eigenvalues calculated on a given gauge configuration which is
$\Z_3$--rotated by $X \in \Z_3$. To be less confusing, for a given
gauge configuration we generate all three $\Z_3$--rotated gauge
configurations and calculate all eigenvalues of the Dirac operator for
each of the three configurations. The Polyakov loop can then be
expressed as a sum over all eigenvalues calculated on all three
rotated gauge configurations, Eq.~\ref{eq:P}. By looking at
Eq.~\ref{eq:P} one immediately may ask what part of the eigenvalue
spectrum contributes most to the Polyakov loop. To answer this
question one introduces the following cumulative sum,
\begin{figure}[htp]
  \centerline{
    \epsfig{file=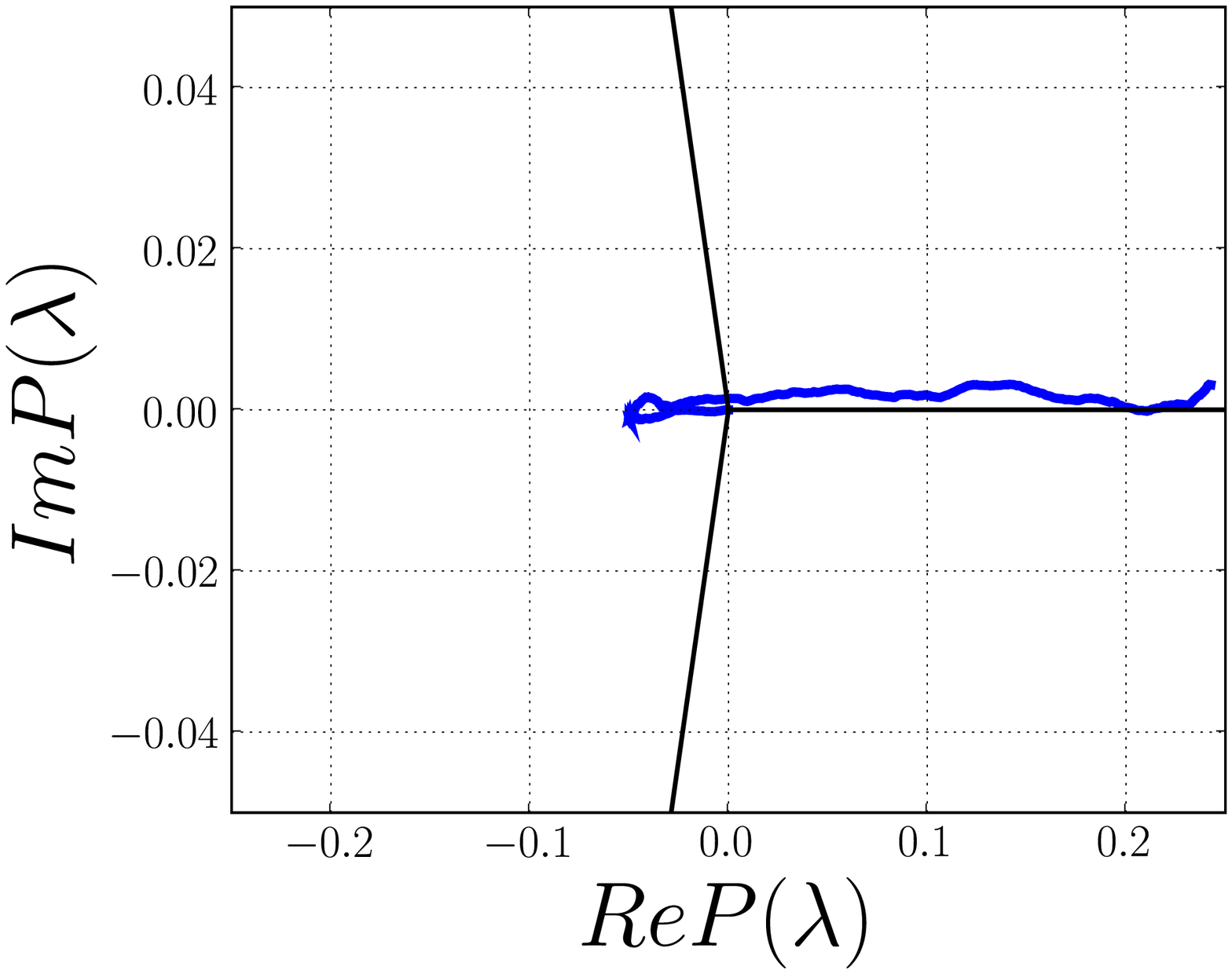, width = 0.5\textwidth}
    \epsfig{file=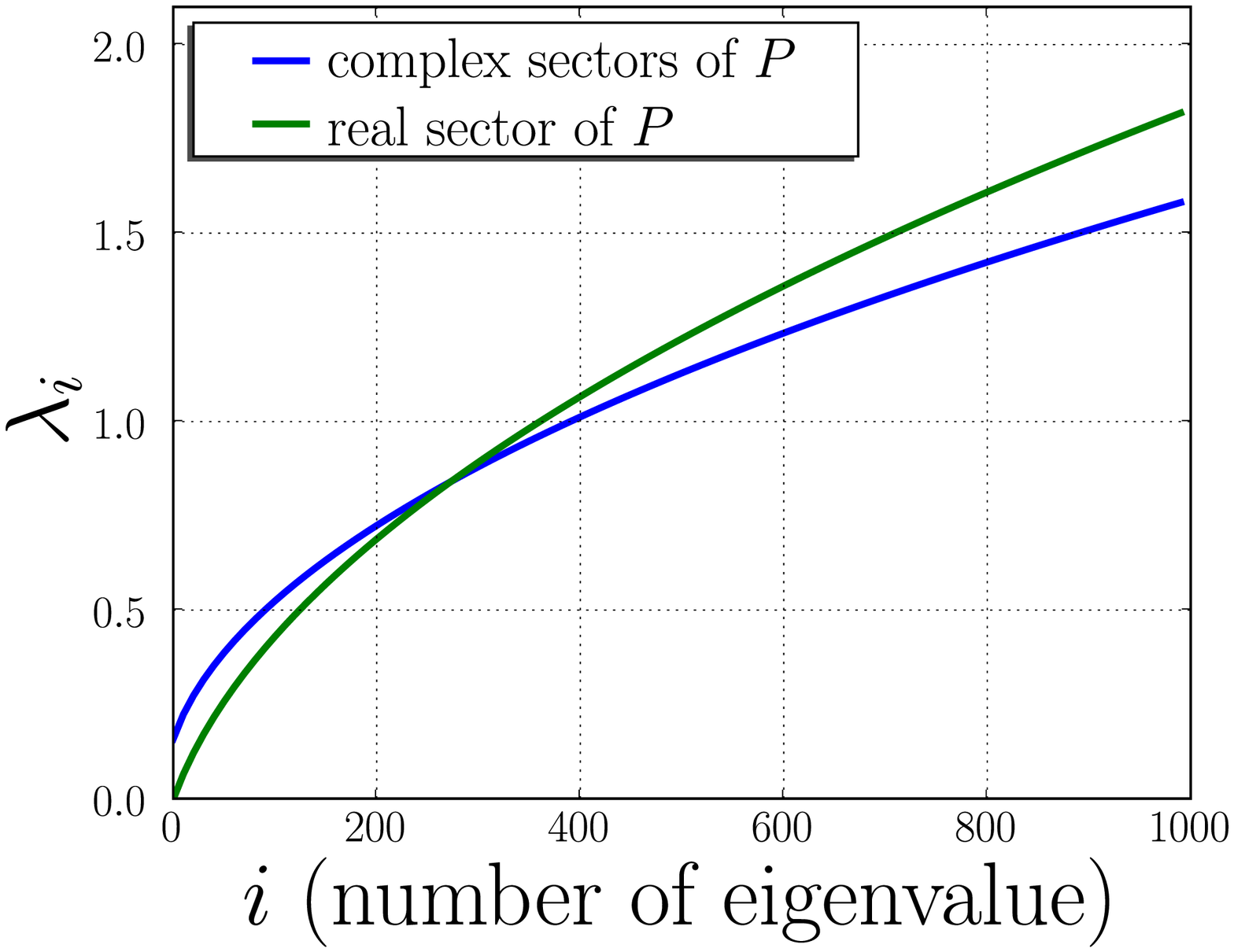, width = 0.5\textwidth}}
  \caption{\label{fig:P} On the left hand side we plotted $P(\lambda)$
    in the complex plane for a typical gauge configuration at $T>T_c$
    for $N_t=4$. The right hand side shows a sketch of the corresponding
    eigenvalues in the real and complex sector of $P$.}
\end{figure}
\begin{equation}
  P(\lambda) = \i^{N_t}\frac{2 \, 2^{N_t}}{3\, N_t\, N_s^3} 
  \sum_{\lambda_{i,X}<\lambda} \left\{1 \cdot \lambda_{i,1}^{N_t} +
    z  \cdot \lambda_{i,z^\ast}^{N_t} + z^\ast  \cdot \lambda_{i,z}^{N_t}
  \right\},
\end{equation}
where we sum over all eigenvalues up to a certain (maximal) value
$\lambda$. Let us briefly make some comments on this formula before
discussing an example. Note that the sector where the Polyakov loop
sits for a given configuration (at $T>T_c$) is solely determined by
multiplying the $\lambda_{i,X}$'s with the appropriate $Z_3$--factors.
Let us assume for the moment that the gauge configuration
corresponding to $\lambda_{i,1}$ has $P$ in the real sector.  Remember
that the two complex sectors of $P$ are physically equivalent.
Therefore, the eigenvalues in the complex sectors of the Polyakov loop
are approximately the same, $\lambda_{i,z} \approx
\lambda_{i,z^\ast}$. Making use of this we obtain for Eq.~\ref{eq:P}
the following approximate relation, $P(\lambda)\sim
\sum_{\lambda_{i,X}<\lambda} ( \lambda_{i,real}^{N_t} -
\lambda_{i,complex}^{N_t} )$. This relation illustrates the fact that
$P(\lambda)$ is built up by the ''response'' of the eigenvalues on the
different $\Z_3$ sectors. As an example we have plotted $P(\lambda)$
for a typical gauge configuration in the complex plane on the left
hand side of figure~\ref{fig:P}. On the right hand side we plotted a
sketch of the corresponding eigenvalues calculated in the real and
complex sector of the Polyakov loop. We observe that for small
eigenvalues the complex sector dominates. By looking at our previously
derived relation we see that this results in negative values of
$P(\lambda)$ while for large eigenvalues it is the other way round.
This shows that the change of the eigenvalues with respect to the
different sectors of the Polyakov loop is crucial.

Finally, we perform the ensemble average on the absolute value of
$P(\lambda)$,
\begin{equation}
  \lk \ls P(\lambda)\rs\rk = \lk \ls \i^{N_t}\frac{2 \,
    2^{N_t}}{3\, N_t\, N_s^3} \sum_{\lambda_{i,X}<\lambda} \left\{1
    \cdot \lambda_{i,1}^{N_t} + z \cdot \lambda_{i,z^\ast}^{N_t} +
    z^\ast \cdot \lambda_{i,z}^{N_t} \right\}\rs\rk.
\end{equation}    
This is the object we will study for the rest of our discussion. At
this point, let us draw the reader's attention to
Refs.~\cite{Bruckmann:2006kx,Synatschke:2007bz,Bilgici:2007} were
similar investigations has been performed using staggered and Wilson
fermions.

\begin{figure}[htp]
  \centerline{
    \epsfig{file=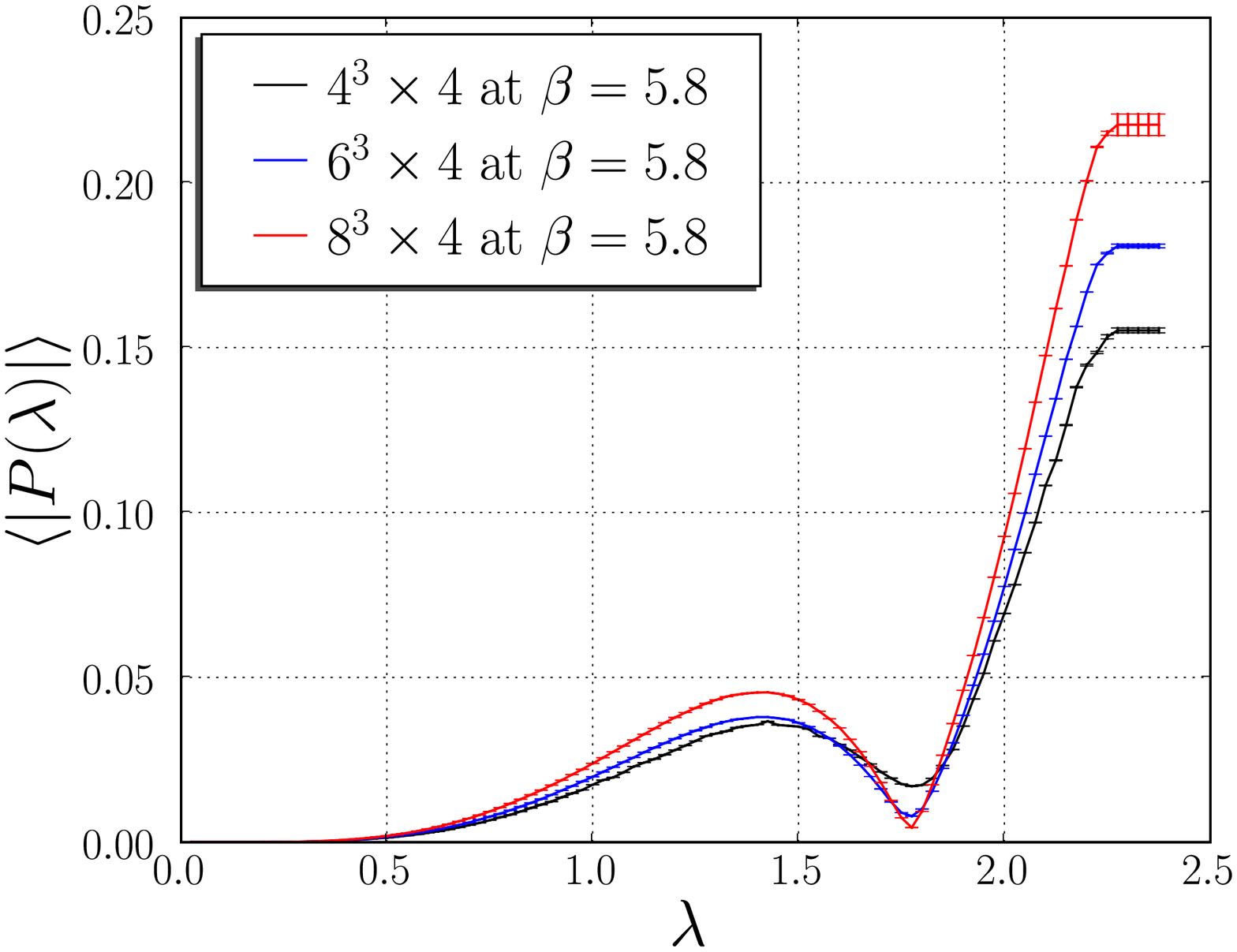, width = 0.5\textwidth}
    \epsfig{file=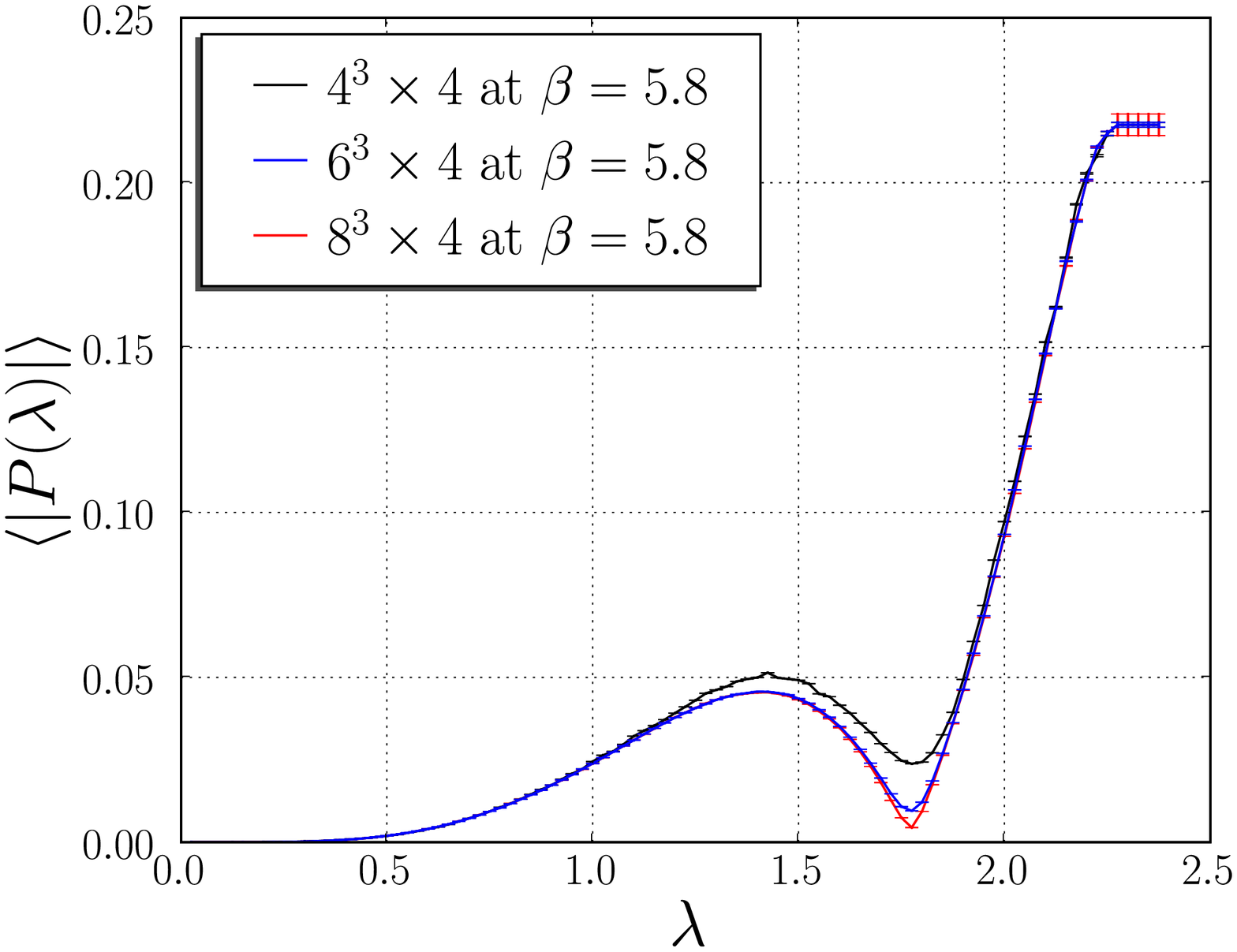, width = 0.5\textwidth}}
  \centerline{
    \epsfig{file=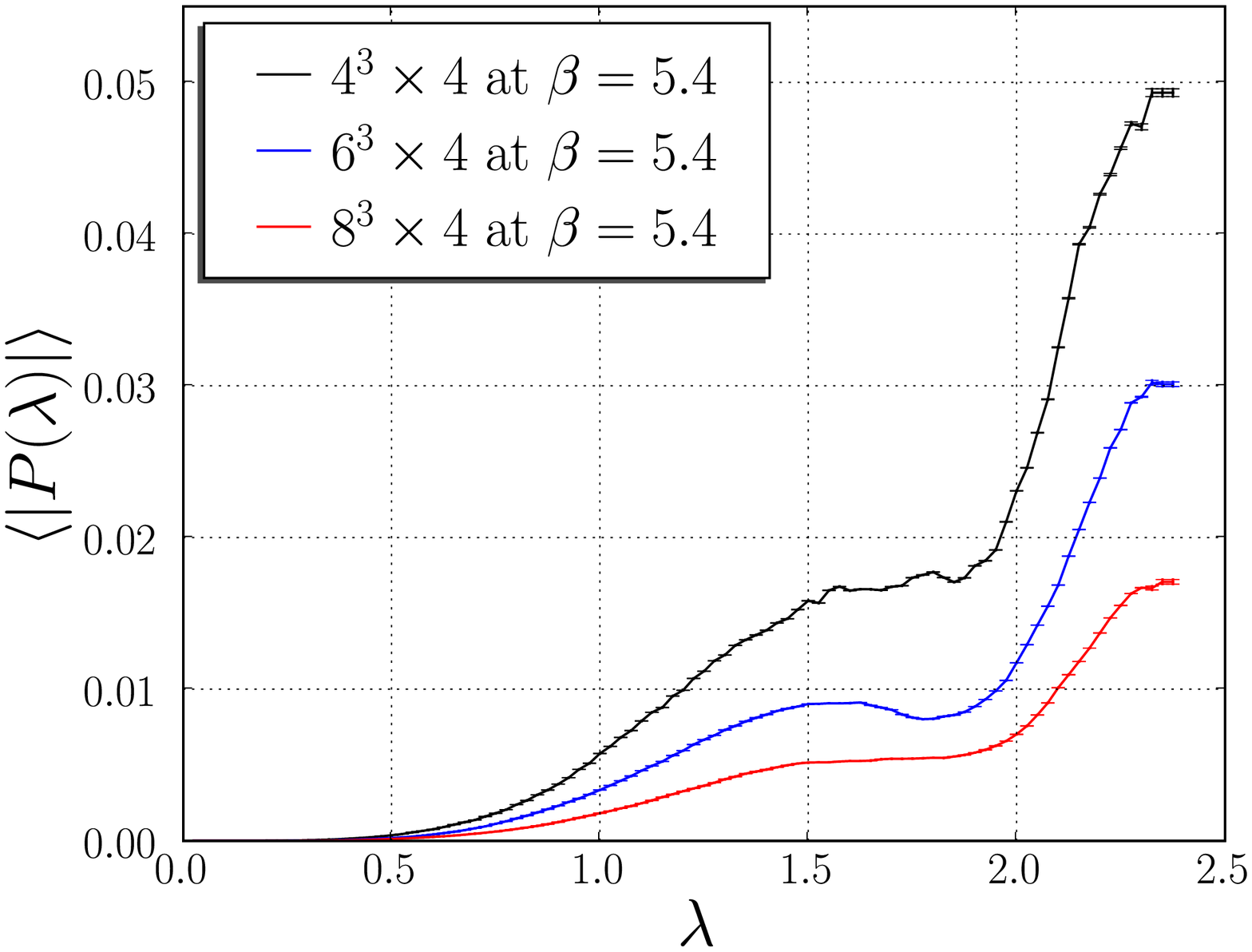, width = 0.5\textwidth}
    \epsfig{file=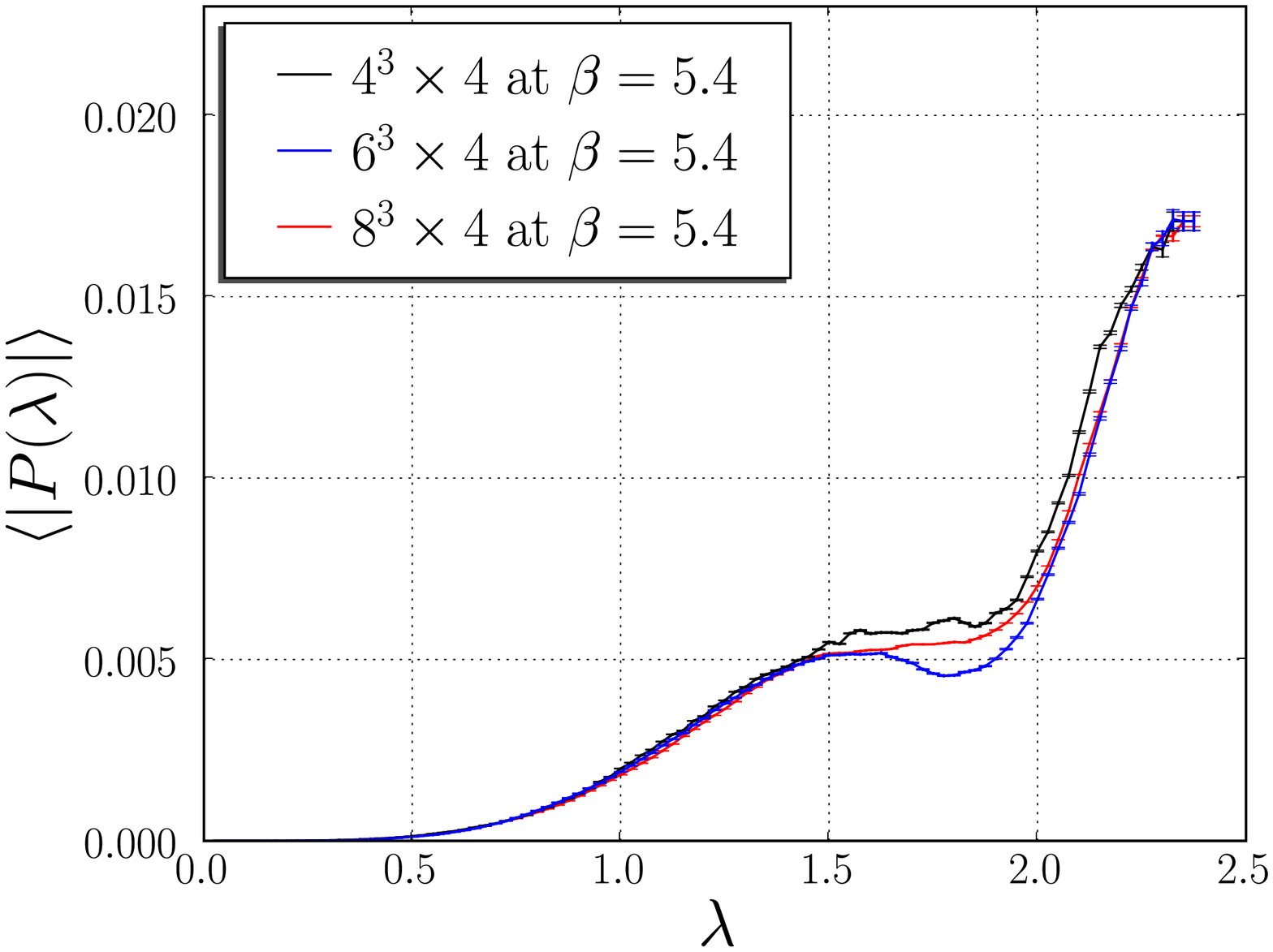, width = 0.5\textwidth}}
  \caption{\label{fig:Pvol}Results for $\lk \ls P(\lambda)\rs\rk$ for
    our quenched configurations. The upper (lower) plots correspond to
    $T>T_c$ ($T<T_c$). The plots on the left hand side illustrate the
    volume dependence, the plots on the right hand side are rescaled
    such that $\lk \ls P(\lambda_{max})\rs\rk$ is the same for all
    three volumes. In the later case we observe that for each $\beta$
    all three curves lie above each other telling us that $\lk \ls
    P(\lambda)\rs\rk$ has the same volume scaling than $P$.}
\end{figure}

\section{Data}
In this section we present several numerical results for $\lk \ls
P(\lambda)\rs\rk$. We start with results for the quenched case where
we have used standard Wilson gauge action. The statistic varies from
$\sim 10$ configurations for the largest lattice $8^4$ up to $\sim
100$ configurations for the smallest lattice. The eigenvalues were
calculated on a single work station using the ARPACK
library~\cite{arpack}.

Let us first take a look at the plot on the upper left side of
figure~\ref{fig:Pvol} where $T>T_c$ and the Polyakov loop is finite.
What we notice right away is that the main contribution to the
Polyakov loop comes from the large eigenvalues which is somewhat
surprising since the physically relevant part of the spectrum should
be the infrared. We will comment on this later on. Another surprising
observation is the dip of the curves at $\lambda \approx 1.7-1.8$.
Naively, one might expect that the cumulative sum $\lk \ls
P(\lambda)\rs\rk$ is a monotonically increasing function. However, by
looking at figure~\ref{fig:P} this behavior becomes clear. The dip in
$\lk \ls P(\lambda)\rs\rk$ (curves in the upper left plot of
figure~\ref{fig:Pvol}) corresponds to the region where $P(\lambda)$
(see figure~\ref{fig:P}) passes zero, the bump in the curves at
$\lambda \approx 1.4$ corresponds to the region where $P(\lambda)$
takes its negative values.  This structure seems to be quite
interesting and one may ask whether it will survive the infinite
volume and the continuum limit.
 \begin{figure}[htp]
  \centerline{
    \epsfig{file=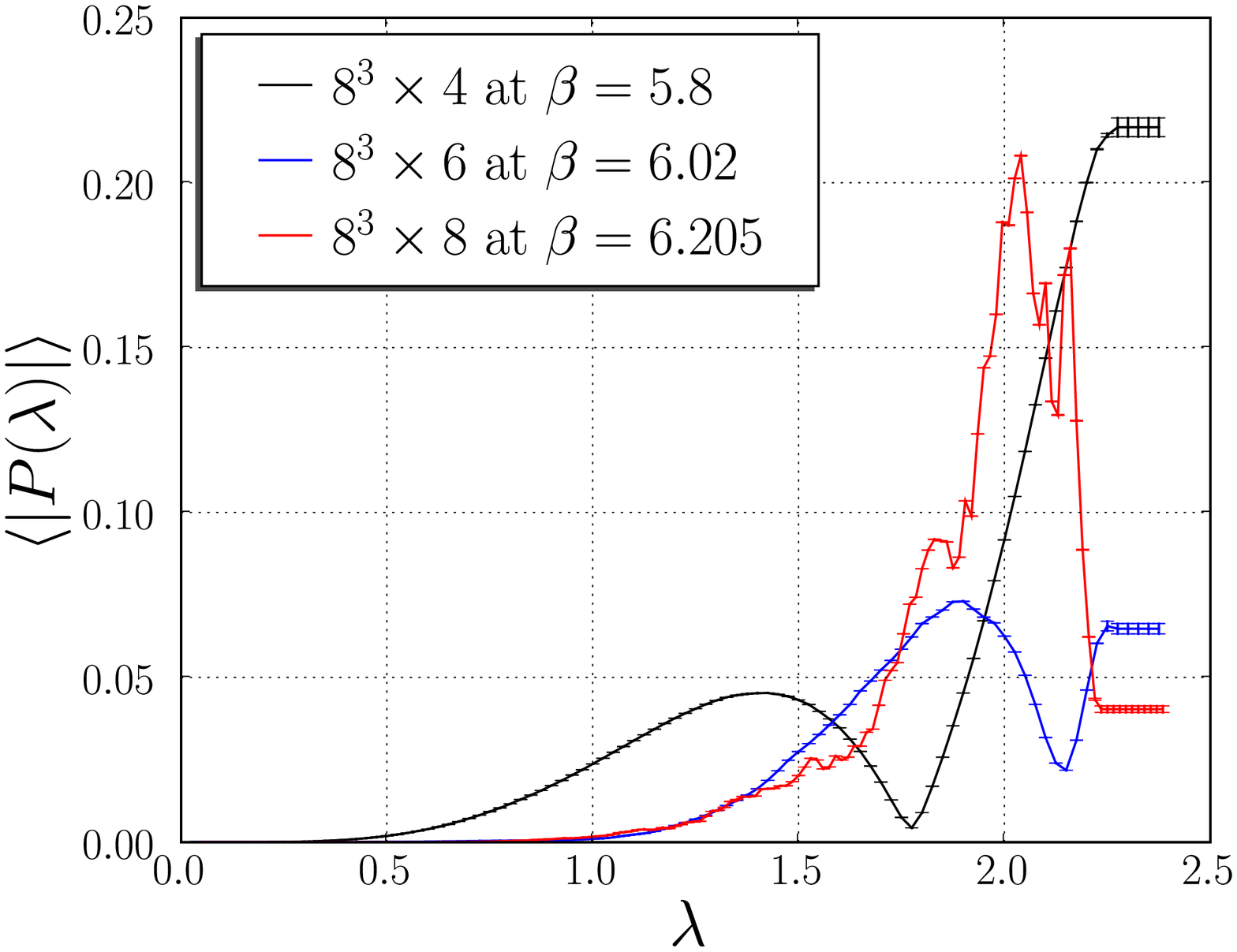, width = 0.5\textwidth}
    \epsfig{file=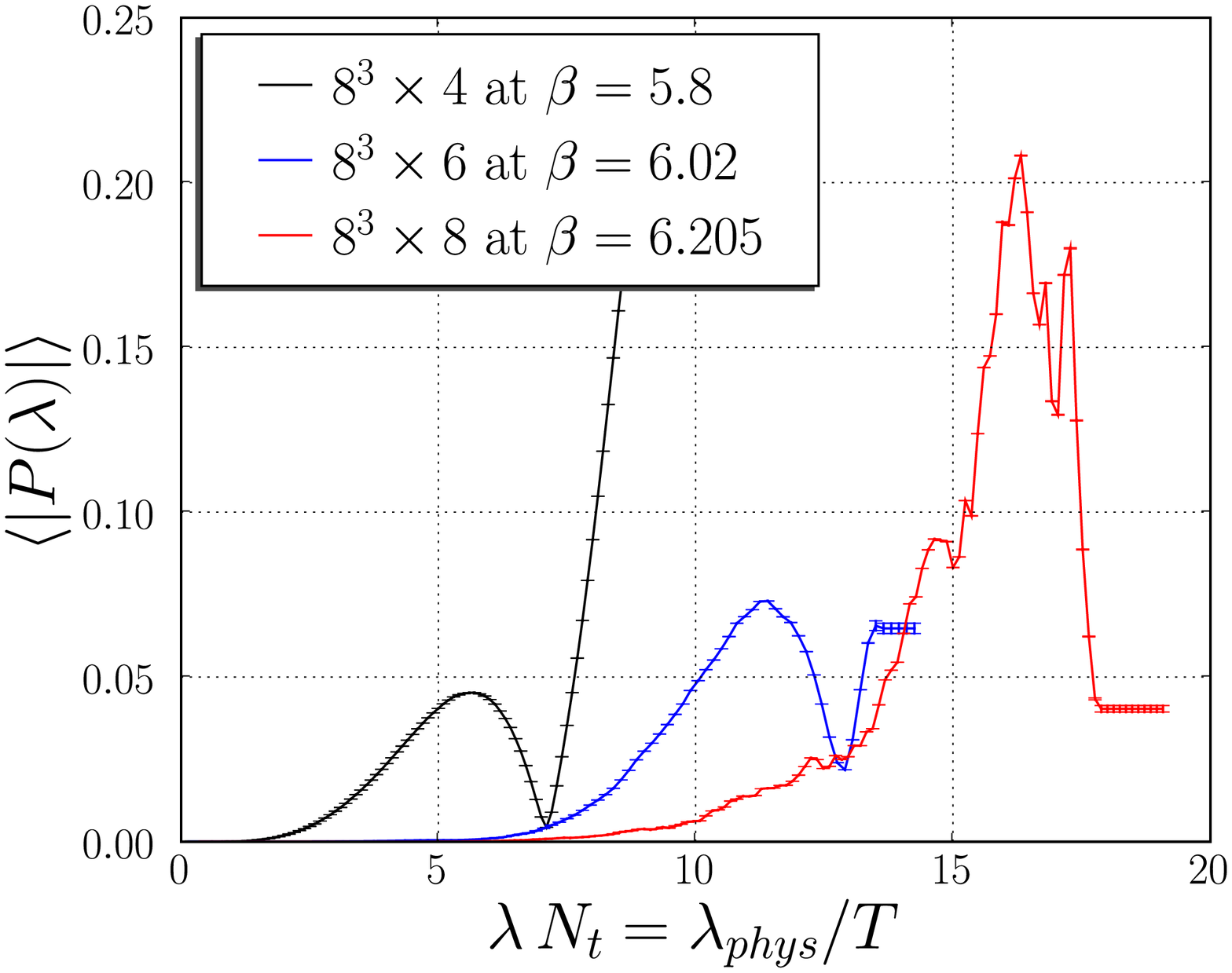, width = 0.5\textwidth}}
  \centerline{
    \epsfig{file=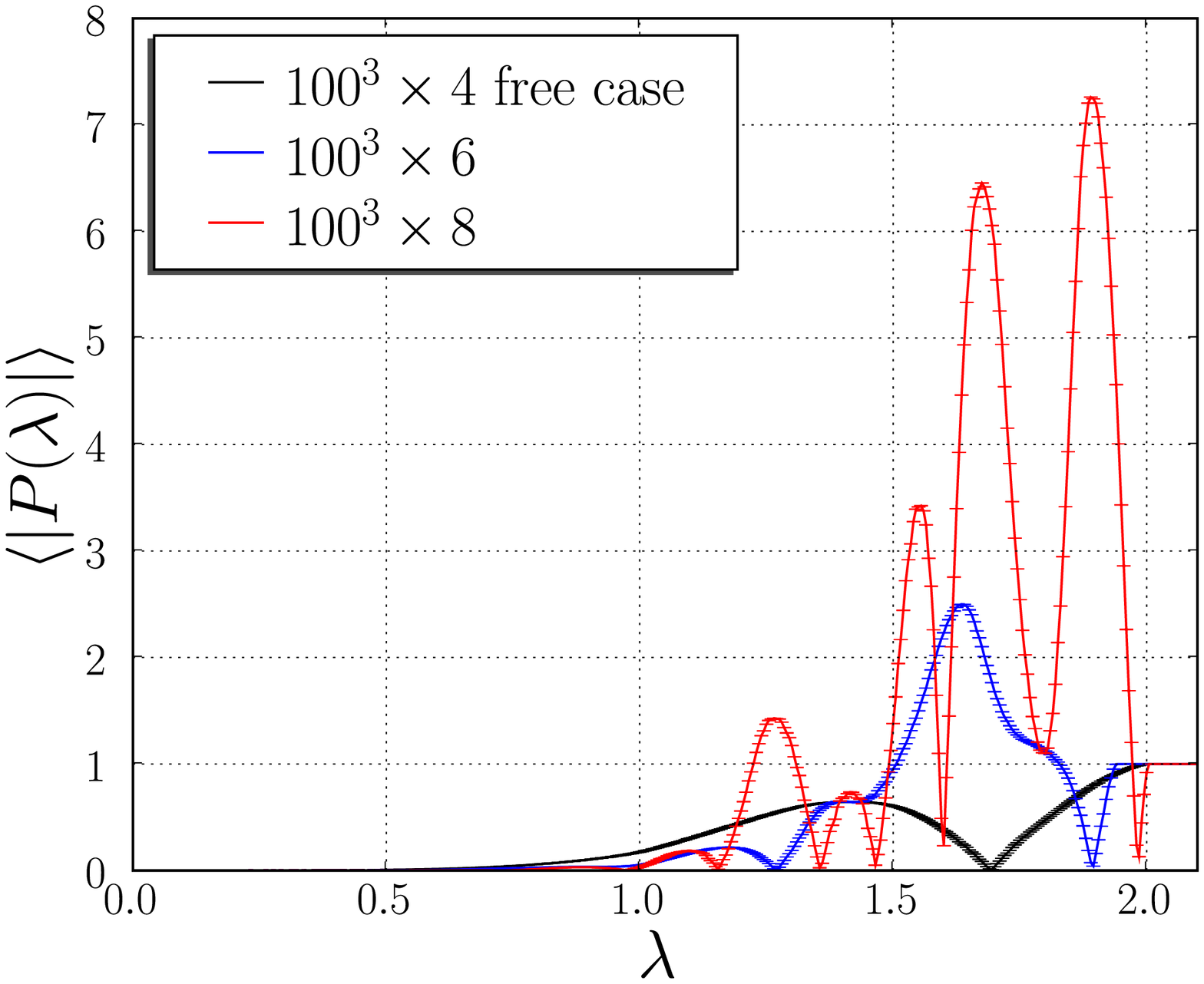, width = 0.5\textwidth}}
  \caption{\label{fig:Pcont}The upper plots show the behavior of $\lk
    \ls P(\lambda)\rs\rk$ as we decrease the lattice spacing $a$ at
    fixed $T \approx 1.2\, T_c$. The plot on the right hand side is
    plotted in physical units. The lower plot shows $\lk \ls
    P(\lambda)\rs\rk$ in the free case for the same values of $N_t$ at
    quite large spatial volume.}
\end{figure}

In figure~\ref{fig:Pvol} we illustrate the volume dependence of $\lk
\ls P(\lambda)\rs\rk$. On the upper plots we show results for three
volumes $4^3$, $6^3$, $8^3$ with $N_t=4$ at a temperature slightly
above $T_c$.  The plot on the upper right hand side shows rescaled
curves where we have fixed $\lk \ls P(\lambda)\rs\rk =\lk \ls
P_{8^3\times 4}\rs\rk$ at $\lambda = \lambda_{max}$ for all three
volumes.  Remember that $P(\lambda_{max})$ is just the ordinary
Polyakov loop $P$.  Beside the curve corresponding to the smallest
volume, which shows small deviations, the curves lie above each other.
This observation tells one that for $T>T_c$ (for sufficiently large
volumes) $\lk \ls P(\lambda)\rs\rk$ has the same volume scaling than
the Polyakov loop itself. This in turn means that the structure will
survive the infinite volume limit.

For $T<T_c$ the situation is similar. On the lower left hand side of
figure~\ref{fig:Pvol} we plotted $\lk \ls P(\lambda)\rs\rk$ for the same
three volumes. The lower right plot shows the corresponding rescaled
curves where we again fixed $\lk \ls P(\lambda_{max})\rs\rk = \lk \ls
P_{8^3\times 4}\rs\rk$. Again, the curves lie above each other showing
that $\lk \ls P(\lambda)\rs\rk$ scales like the Polyakov loop also
below $T_c$.  Because below $T_c$ the Polyakov loop vanishes in the
infinite volume limit we find that also $\lk \ls P(\lambda)\rs\rk$
will vanish in this limit. So in this case the structure does not
survive the infinite volume limit. Note that knowing the volume
dependence of $\lk \ls P(\lambda)\rs\rk$ will keep the computational
costs significantly lower because one do not have to perform expensive
computations on large volume lattices.
\begin{figure}[htp]
  \centerline{
    \epsfig{file=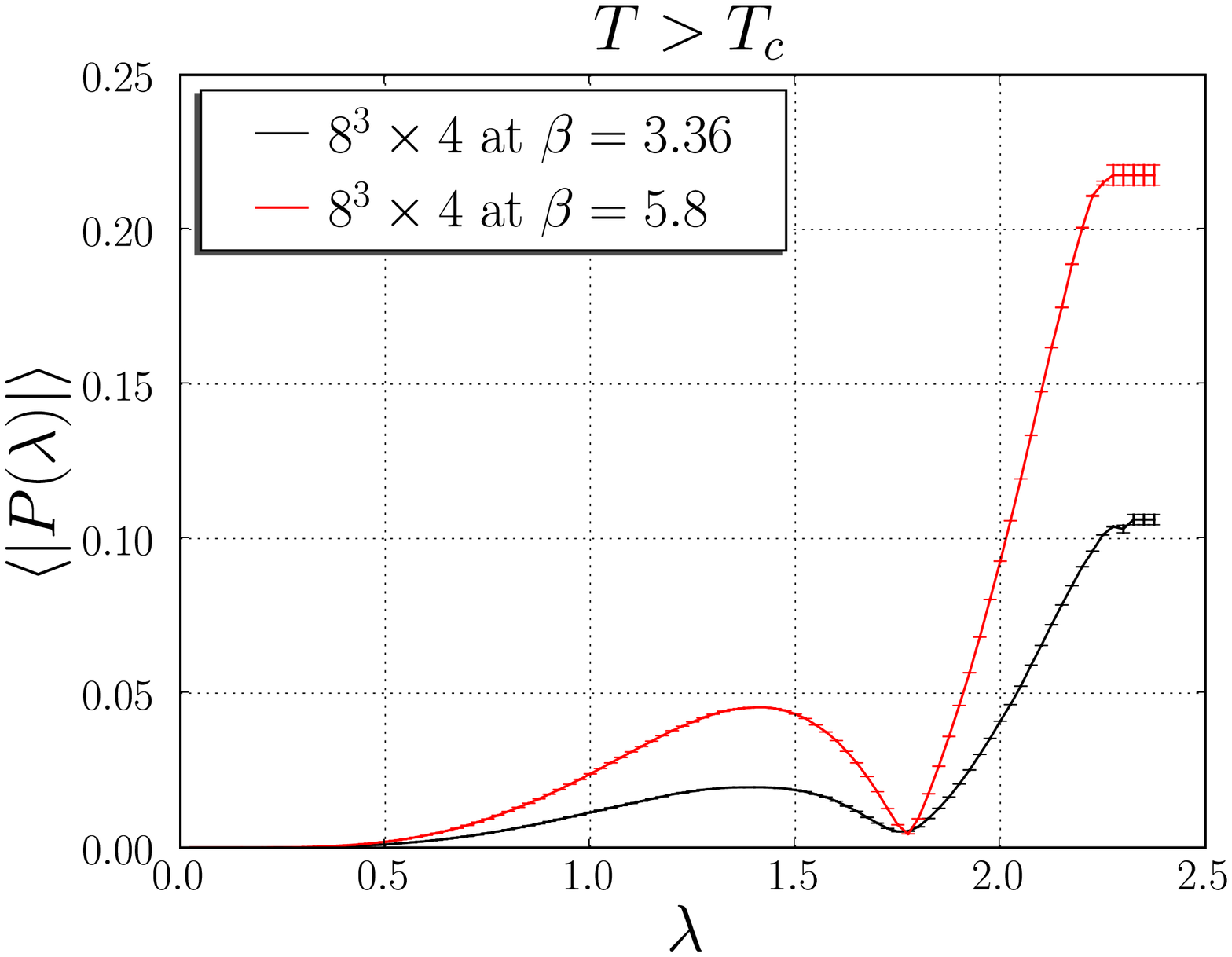, width = 0.50\textwidth}
    \epsfig{file=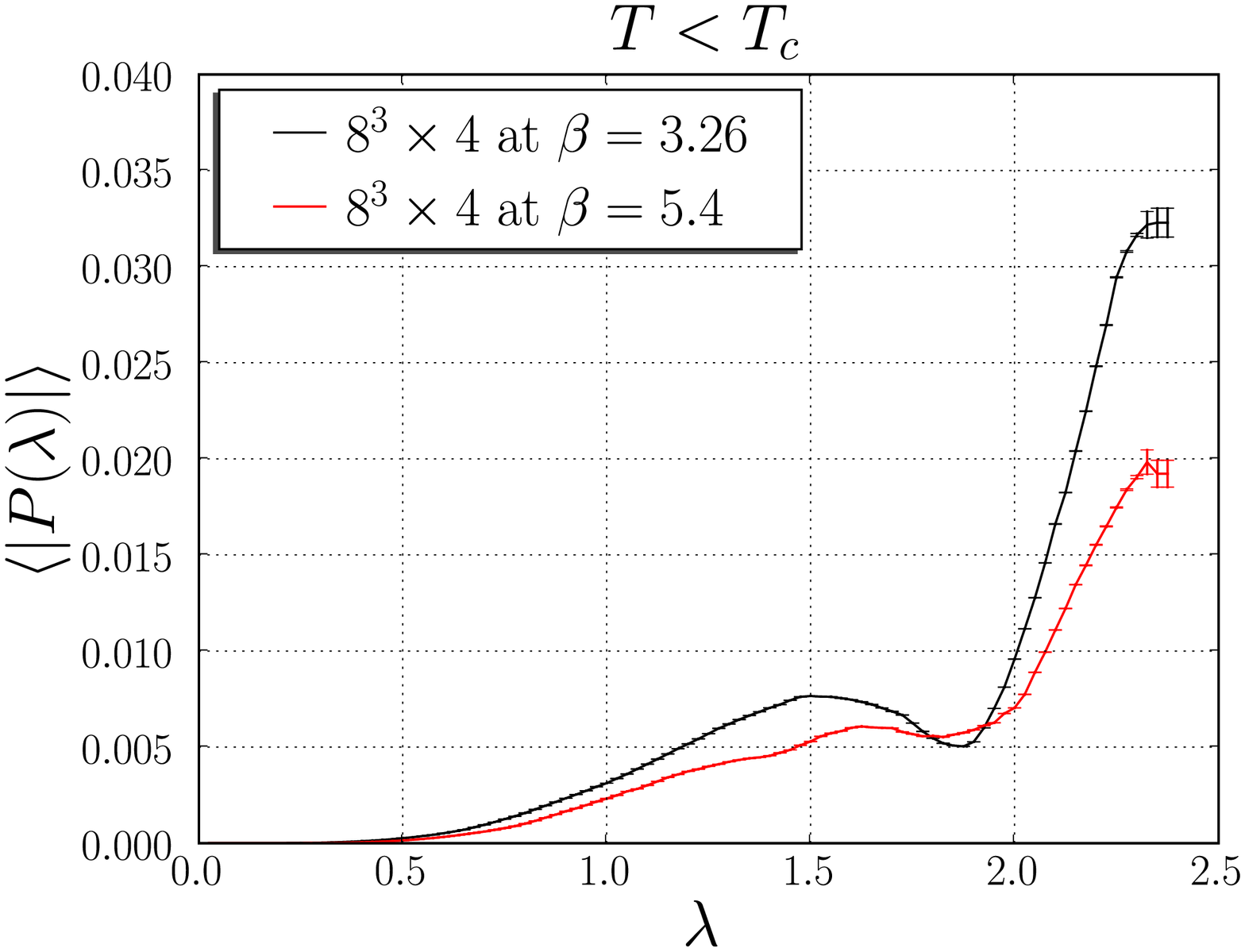, width = 0.50\textwidth}}
  \caption{\label{fig:Pdyn}We plotted $\lk \ls P(\lambda)\rs\rk$ using
    dynamical ($\beta=3.26,3.36$) and quenched ($\beta=5.4,5.8$)
    configurations. The dynamical configurations were generated with
    masses $m_q$=0.0065 and $m_s$=0.065. The temperature in the
    dynamical case differs slightly from the quenched one.}
\end{figure}

Let us now look at how $\lk \ls P(\lambda)\rs\rk$ behaves as the
lattice spacing $a \to 0$. Since for $T<T_c$ $\lk \ls
P(\lambda)\rs\rk$ vanishes in the infinite volume limit anyway we will
discuss only results for $T>T_c$. The upper plots in
figure~\ref{fig:Pcont} illustrate our results for $N_t=4,6,8$ at fixed
$T \approx 1.2 \, T_c$ plotted against $\lambda$ in lattice and
physical units. Let us focus on the upper left plot. We find that for
$N_t=6$ there is a bump and a dip in the curve similar to that in the
curve for $N_t=4$. For $N_t=8$ we also observe a large bump but
because in this case the statistic is quite limited and the spatial
volume is rather small the signal is quite noisy. We notice that, as
we go to smaller lattice spacing, the structures in the curves move
towards the ultraviolet (UV) part of the eigenvalue spectrum and the
Polyakov loop obtains its final value more from the very end of the UV
part of the spectrum.

Let us compare these results to the free case where an analytical
expression for the eigenvalues is known. The lower plot in
figure~\ref{fig:Pcont} shows our results in the free case for the same
three values of $N_t$ at quite large spatial volume. We find that the
shape as well as the position of the bumps and dips of the curves are
surprisingly similar to the corresponding results of the quenched data
(upper left plot). This leads to the following conclusion. As we
approach the continuum limit $\lk \ls P(\lambda)\rs\rk$ at small
values of $\lambda$ is essentially zero. At somewhat large eigenvalues
$\lk \ls P(\lambda)\rs\rk$ starts to show wild fluctuations which
cancel out at the very end of the UV part of the eigenvalue spectrum
where the Polyakov loop obtains its final value.

We remark that it might be not too surprising that the Polyakov loop
is dominated by the UV part of the eigenvalue spectrum as the
Polyakov loop is related to the propagation of an infinitely heavy
quark. By looking at the quark propagator in the spectral
representation,
\begin{equation}
  S(x,y)=\sum_\lambda \frac{\psi_\lambda(x)
    \psi_\lambda^\dagger(y)}{\lambda +\i m},  
\label{eq:qp}
\end{equation}
where $\psi_\lambda(x)$ are the normalized eigenvectors of the Dirac
operator, we note that the eigenmodes under the sum are weighted by
$(\lambda +\i m)^{-1}$. For a very heavy quark (as $m \to \infty$) the
relative weight of each eigenmode becomes approximately the same.
Therefore, UV eigenmodes can dominate the propagation of an
infinitely heavy quark, i.e.~the Polyakov loop.

Finally, we compare the quenched results to dynamical results using
$p4fat3$ fermions with improved gauge action and quark masses
$m_q$=0.0065 and $m_s$=0.065, see Ref.~\cite{p4fat3}. On the left
(right) hand side of figure~\ref{fig:Pdyn} we plotted $\lk \ls
P(\lambda)\rs\rk$ for $T>T_c$ ($T<T_c$). Surprisingly, there is no
qualitative difference in the behavior of $\lk \ls P(\lambda)\rs\rk$
in the dynamical case.

\section{Summary}

In this work we have studied the connection between the Polyakov loop
and the eigenvalues of the Dirac operator using $\lk \ls
P(\lambda)\rs\rk$. We have focused on the volume dependence and the
continuum limit. We have found that the dominant contribution to the
Polyakov loop comes from the very end of the UV part of the eigenvalue
spectrum. We also compared our results to the free case. A comparison
between full QCD and quenched QCD seems to show no qualitative
difference.

Our findings suggest that the dependence of the eigenvalues on the
different $\Z_3$ sectors of the Polyakov loop seems to be crucial.
Aiming at the connection between confinement and chiral symmetry
breaking our findings can be concluded in the following picture. Above
$T_c$ the eigenvalues $\lambda$ show a strong dependence on the
different sectors of the Polyakov loop resulting in wild fluctuations
in $\lk \ls P(\lambda)\rs\rk$. These fluctuations cancel out in a way
that the Polyakov loop obtains its finite value from the very end of
the UV spectrum. At the same time, since the chiral condensate is zero
above $T_c$, the infrared (IR) part of the spectrum shows a vanishing
density of eigenvalues.

Below $T_c$ the dependence of the eigenvalues on the Polyakov loop
sectors vanishes in the infinite volume limit which leads to a
vanishing $\lk \ls P(\lambda)\rs\rk$. In particular $\lk \ls
P(\lambda)\rs\rk$ vanishes at the UV leading to a Polyakov loop $\lk
\ls P\rs\rk=0$. At the same time, the eigenvalue density at the IR
part of the spectrum becomes finite since chiral symmetry is broken.

\acknowledgments The author would like to thank Erek Bilgici, Falk
Bruckmann, Christof Gattringer, Kay H\"ubner, Christian Hagen,
Frithjof Karsch, and Christian Schmidt for stimulating discussions.

\appendix

\end{document}